\documentclass[%
  aps,
  prx,
  reprint,
  superscriptaddress,
  longbibliography, 
  amsfonts,
  amssymb,
  amsmath,
  floatfix, 
  eprint, 
  letterpaper,
  citeautoscript, 
]{revtex4-2}

\pdfoutput=1

\usepackage[utf8]{inputenc}

\usepackage[english]{babel}

\usepackage{dsfont}                 
\usepackage{bbold}                  

\usepackage[rgb,dvipsnames]{xcolor} 

\usepackage{comment}                
\usepackage{todonotes}              
\setuptodonotes{inline,
                backgroundcolor=yellow!30,
                prepend,
                caption={\tt}}

\usepackage[en-US]{datetime2}

\usepackage{graphicx}                

\usepackage{float}                   
\usepackage[export]{adjustbox}

\usepackage{tabulary}         
\usepackage{multirow}

\usepackage{mathtools}         
\usepackage{algorithm}        
\usepackage{algorithmic}

\usepackage[colorlinks=true,
            urlcolor=blue,
            citecolor=blue,
            linkcolor=blue]{hyperref}

\usepackage{qcircuit}
\usepackage{braket}

\usepackage{silence}
\graphicspath{{figures/}}

\definecolor{mygreen}{RGB}{34, 136, 51}
\definecolor{myblue}{RGB}{68, 119, 170}
\definecolor{myred}{RGB}{238, 102, 119}
\definecolor{mypurple}{RGB}{170, 51, 119}
\definecolor{mycyan}{RGB}{102, 204, 238}

\renewcommand{\>}{\rangle}

\newcommand{\lsim}{\mathrel{\raise.3ex\hbox{$<$\kern-.75em\lower1ex\hbox{$\sim$}}}}
\newcommand{\gsim}{\mathrel{\raise.3ex\hbox{$>$\kern-.75em\lower1ex\hbox{$\sim$}}}}

\def\QECCnk[[#1,#2]]{[\![#1, #2]\!]}

\def\QECCnkq[[#1,#2,#3]]{[\![#1, #2]\!]_{#3}^{\vphantom{T}}}

\def\QECCnkd[[#1,#2,#3]]{[\![#1, #2, #3]\!]}

\def\QECCnkdq[[#1,#2,#3,#4]]{[\![#1, #2, #3]\!]_{#4}^{\vphantom{T}}}

\def\QECCnkgd[[#1,#2,#3,#4]]{[\![#1, #2, #3, #4]\!]}

\def\QECCnkgdq[[#1,#2,#3,#4,#5]]{[\![#1, #2, #3, #4]\!]_{#5}^{\vphantom{T}}}

\def\QECCnkdc[[#1,#2,#3,#4]]{[\![#1, #2, #3; #4]\!]}

\def\QECCnkdcq[[#1,#2,#3,#4,#5]]{[\![#1, #2, #3; #4]\!]_{#5}^{\vphantom{T}}}

\def\QECCnkgdcq[[#1,#2,#3,#4,#5,#6]]{%
  [\![#1, #2, #3, #4; #5]\!]_{#6}^{\vphantom{T}}}

\newcommand{\bigO}{{\cal O}}

\newcommand\CNOT{\ensuremath{\textit{CNOT\/}}}

\newcommand\TOF{\ensuremath{\textit{TOF\/}}}

\def\openone{\leavevmode\hbox{\small1\normalsize\kern-.33em1}}

\newcommand{\etal}{\textit{et~al.}}
\newcommand{\viz}{\textit{viz.}}

\newcommand{\eg}{\textit{e.g.}}

\long\def\symbolfootnote[#1]#2{\begingroup%
\def\thefootnote{\fnsymbol{footnote}}\footnote[#1]{#2}\endgroup}

\newlength{\blob}
\newcommand{\sx}[1]{\makebox[\blob][c]{$#1$}}

\newcommand{\cubecode}{\ensuremath{[\![8, 3, 2]\!]}}
\newcommand{\squarecode}{\ensuremath{[\![4, 2, 2]\!]}}
\newcommand\CCZ{\ensuremath{\textit{CCZ\/}}}

\begin{document}

\title{Below-threshold execution of fault-tolerant ripple-carry addition on a trapped-ion quantum computer}

\author{Jacob S. Nelson}
\email{jnelso2@sandia.gov}
\affiliation{Quantum New Mexico Institute (QNM-I),
             Department of Physics and Astronomy,
             University of New Mexico,
             Albuquerque, NM 87131, USA}
\affiliation{Quantum Algorithms and Applications Collaboratory (QuAAC),\\           
             Sandia National Laboratories,
             Albuquerque, NM 87185, USA}

\author{Andrew J. Landahl}
\email{alandahl@sandia.gov}
\affiliation{Quantum New Mexico Institute (QNM-I),
             Department of Physics and Astronomy,
             University of New Mexico,
             Albuquerque, NM 87131, USA}
\affiliation{ Microsystems Engineering, Systems and Applications (MESA),
             Quantum Architectures Group (Q-Arch),\\      
             Sandia National Laboratories,
             Albuquerque, NM 87185, USA}

\author{Andrew D. Baczewski}
\email{adbacze@sandia.gov}
\affiliation{Quantum New Mexico Institute (QNM-I),
             Department of Physics and Astronomy,
             University of New Mexico,
             Albuquerque, NM 87131, USA}
\affiliation{Quantum Algorithms and Applications Collaboratory (QuAAC),\\
             Sandia National Laboratories,
             Albuquerque, NM 87185, USA}

\begin{abstract}

We deploy a universal fault-tolerant quantum computing (FTQC) architecture based on error-detecting \cubecode\ and \squarecode\ codes on Quantinuum’s Helios-1 trapped-ion quantum computer.
We compile and execute ripple-carry adder (RCA) circuits for classical inputs up to five bits on this architecture.
Uncomputing a workspace qubit provides an additional ``algorithmic check'' to detect errors.
While available at both the physical and logical levels, this check synergizes with our architecture by increasing the effective fault distance of many circuit locations from two to four.
Combining this algorithmic check with the error-detecting codes, we observed only a single logical failure in one-bit addition after burst-error filtering, yielding an estimated logical failure rate of $(6.0 \pm 5.9) \times 10^{-5}$.
This is well below the $(3.5 \pm 1.3) \times 10^{-3}$ failure probability observed for one-bit addition on the same hardware in unencoded form.
Two- and three-bit adders also showed performance improvements, while the longer four- and five-bit adders executed just above threshold.

\end{abstract}

\date{\DTMnow}

\maketitle

\section{Introduction}

Quantum hardware has progressed beyond the noisy intermediate-scale quantum (NISQ) era \cite{Preskill:2018b} and into the early fault-tolerant quantum computing (EFTQC) era in which fault-tolerant protocols are used to suppress error spreading, but at scales too small to implement interesting quantum algorithms~\cite{preskill2025beyond}. 
This transition is heralded by experiments demonstrating the encoding and manipulation of small-to-moderate numbers of logical qubits~\cite{ryan2021realization, acharya2024quantumerrorcorrectionsurface, Bluvstein_2023, pogorelov2025experimental, ryan2024high, reichardt2024demonstrationquantumcomputationerror, reichardt2024logical, lacroix2025scaling, eickbusch2025demonstration, dasu2025breaking, rosenfeld2025magic, berthusen2026experimental}.
Recent work has demonstrated fault-tolerant algorithm execution~\cite{Wang_2024, butt2025demonstrationmeasurementfreeuniversalfaulttolerant, perlin2026fault, wang2026superconductingsurfacecodeprocessorlatticesurgery}, with results operating near or slightly below the error thresholds needed for these protocols to outperform their physical counterparts.
Here, we report the compilation of a family of ripple-carry adders (RCA)~\cite{cuccaro2004newquantumripplecarryaddition} into an EFTQC architecture based on \cubecode\ and \squarecode\ quantum-error correcting (QEC) codes that we developed in earlier work~\cite{nelson2025smallinterestingarchitectureearly}.
We then compare the physical and logical execution of these RCA circuits on instances up to five bits on Quantinuum's Helios-1, a 98-qubit trapped-ion platform~\cite{ransford202698}.

Arithmetic circuits are foundational building blocks in many quantum algorithms~\cite{gidney2018halving, babbush_2018, rubin2024quantum, gidney2025factor, pathak2026approach}, and thus important demonstrators for EFTQC architectures.
Indeed, one of the first experimental demonstrations in the EFTQC era implemented a one-bit half adder on Quantinuum's H1-1 20-qubit platform using the same \cubecode\ code used in our architecture~\cite{Wang_2024}.
We build on this demonstration by applying it to full adders that propagate carry bits on inputs up to five bits.
We further demonstrate that checking that the restoration of the initial $|0\rangle$ carry qubit that seeds the algorithm greatly enhances error suppression in the adders.  
This ``algorithmic check'' increases many RCA circuit locations from a fault distance of two to a fault distance of four.  
In other words, the redundant ``gauge'' degree of freedom in the algorithm works with underlying QEC codes to suppress errors even more than the codes could on their own.

The rest of the paper is organized as follows. 
In Section~\ref{sec:SIA}, we review our architecture.
In Section~\ref{sec:RCA}, we compile a well-studied RCA algorithm onto this architecture and discuss how the algorithmic check interacts with the associated QEC codes.
In Section~\ref{sec:results}, we present our experimental results.
Finally, in Section~\ref{sec:conclusion}, we summarize our results and discuss possible extensions of this work.

\section{Review of our architecture}
\label{sec:SIA}

In Ref.~\cite{nelson2025smallinterestingarchitectureearly}, we presented the EFTQC architecture that we use in this work.
We briefly review it here, for completeness.

\subsection{Universal gate set}
\label{sec:universal_gate_set}

Our architecture's universal gate set consists of the controlled-controlled-$Z$ ($\CCZ$) gate and the Hadamard ($H$) gate, along with the ability to prepare and measure qubits in the $Z$ basis~\cite{Shi:2003a, aharonov2003simple}.  
In order to represent complex numbers, it uses an additional phase-reference qubit to track the real and imaginary components of states and gates, such as the gate $T = Z^{1/4}$.
However, because the RCA algorithm only uses real-valued gates, we restrict our attention to quantum states with real coefficients to avoid the overhead of this encoding. 
It is straightforward to re-introduce the phase-reference qubit to implement the RCA algorithm for complex superpositions of addends, if desired.

\begin{figure}[!h]
    \centering
    \includegraphics[width=0.3\textwidth]{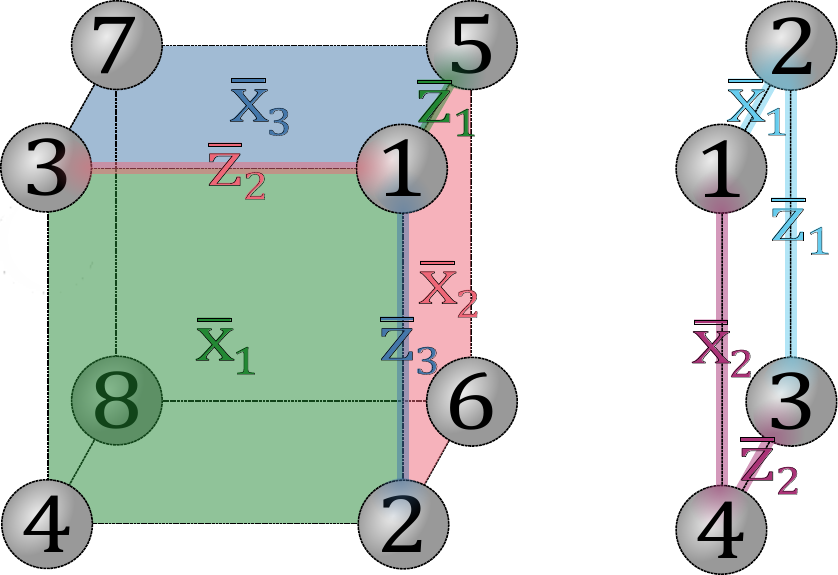}
    \caption{Encoded logical operators of the \cubecode\ code (left) and the \squarecode\ code (right). Circles depict physical qubits and the colored planes/edges depict logical operators.}
    \label{fig:cube-square}
\end{figure}

The codes used by our architecture are the \cubecode\ code, dubbed the ``smallest interesting colour code'' by Earl Campbell~\cite{Campbell:2016a}, and a \squarecode\ code, dubbed an ``iceberg'' code by Self \etal~\cite{Gottesman:1998a, Self:2024a}. 
They are depicted graphically in Fig.~\ref{fig:cube-square}, and algebraically in Table~\ref{tab:codes}.
The \cubecode\ code realizes the logical \CCZ\ gate by transversal single-qubit $T/T^\dagger$ gates and the \squarecode\ code realizes the logical $(H\otimes H)\textit{SWAP}$ gate by transversal single-qubit $H$ gates.
Furthermore, both are examples of ``phantom codes''~\cite{koh2026entanglinglogicalqubitsphysical}, for which intra-block $\CNOT$\ gates can be enacted by passive physical qubit relabelings, rendering these logical gates essentially free.

\begin{table}[!h]
 \centering
 \setlength{\extrarowheight}{5pt} 
    \begin{tabular}{|c|c|c|c|}
        \hline
        & \raisebox{0.235em}{\cubecode\ }  & \raisebox{0.235em}{\squarecode\ } \\
        \hline
        \parbox[t]{3mm}{\multirow{5}{*}{\rotatebox[origin=c]{90}{Stabilizers}}} & $S_1 = 
         \sx{Z}\sx{Z}\sx{Z}\sx{Z}\sx{I}\sx{I}\sx{I}\sx{I}$ & $S_1 = \sx{X}\sx{X}\sx{X}\sx{X}$
         \\
         & $S_2 = \sx{Z}\sx{Z}\sx{I}\sx{I}\sx{Z}\sx{Z}\sx{I}\sx{I}$ & $S_2 = \sx{Z}\sx{Z}\sx{Z}\sx{Z}$ \\ 
         & $S_3 = \sx{Z}\sx{I}\sx{Z}\sx{I}\sx{Z}\sx{I}\sx{Z}\sx{I}$ & \\
         & $S_4 = \sx{Z}\sx{Z}\sx{Z}\sx{Z}\sx{Z}\sx{Z}\sx{Z}\sx{Z}$ & \\
         & $S_5 = \sx{X}\sx{X}\sx{X}\sx{X}\sx{X}\sx{X}\sx{X}\sx{X}$ &  \\
         \hline 
         \parbox[t]{3mm}{\multirow{6}{*}{\rotatebox[origin=c]{90}{Logical Operators}}}
         & \textcolor{mygreen}{$\overline{X}_1 = \sx{X}\sx{X}\sx{X}\sx{X}\sx{I}\sx{I}\sx{I}\sx{I}$} & \textcolor{mycyan}{$\overline{X}_1 = \sx{X}\sx{X}\sx{I}\sx{I}$} \\ 
         & \textcolor{mygreen}{$\overline{Z}_1 = \sx{Z}\sx{I}\sx{I}\sx{I}\sx{Z}\sx{I}\sx{I}\sx{I}$} & \textcolor{mycyan}{$\overline{Z}_1 = \sx{I}\sx{Z}\sx{Z}\sx{I}$} \\
         & \textcolor{myred}{$\overline{X}_2 = \sx{X}\sx{X}\sx{I}\sx{I}\sx{X}\sx{X}\sx{I}\sx{I}$} & \textcolor{mypurple}{$\overline{X}_2 = \sx{X}\sx{I}\sx{I}\sx{X}$} \\
         & \textcolor{myred}{$\overline{Z}_2 = \sx{Z}\sx{I}\sx{Z}\sx{I}\sx{I}\sx{I}\sx{I}\sx{I}$} & \textcolor{mypurple}{$\overline{Z}_2 = \sx{I}\sx{I}\sx{Z}\sx{Z}$} \\
         & \textcolor{myblue}{$\overline{X}_3 = \sx{X}\sx{I}\sx{X}\sx{I}\sx{X}\sx{I}\sx{X}\sx{I}$} & \\
         & \textcolor{myblue}{$\overline{Z}_3 = \sx{Z}\sx{Z}\sx{I}\sx{I}\sx{I}\sx{I}\sx{I}\sx{I}$} & \\
 \hline
 \end{tabular}
 \caption{Algebraic representation of the stabilizer generators and logical operators of the \cubecode\ (left) and \squarecode\ (right) codes. The color of the text depicting the logical operators is chosen to match the corresponding colored edges/faces in the graphical representation displayed in Fig.~\ref{fig:cube-square}.}
 \label{tab:codes}
 \end{table}

\begin{figure}[!h]
    \centering
    \includegraphics{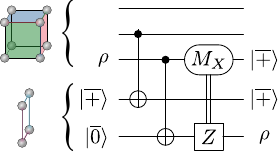}
    \caption{Logical teleportation circuits showing  teleportation of a single state from a \cubecode\ code to a \squarecode\ using the transversal $\CNOT^{\otimes2}$ gate between code blocks. 
    $M_{X}$ denotes a projective measurement onto the $+1$ eigenstate of the measured operator. }
    \label{fig:X_teleport_circ}
\end{figure}

\begin{figure}[!h]
    \centering
    \includegraphics{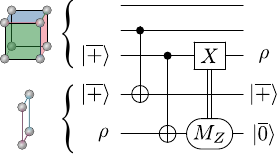}
    \caption{Logical teleportation circuits showing  teleportation of a single state from a \squarecode\ code to a \cubecode\ code using the transversal $\CNOT^{\otimes2}$ gate between code blocks. 
    $M_{Z}$ denotes a projective measurement onto the $+1$ eigenstate of the measured operator.}
    \label{fig:Z_teleport_circ}
\end{figure}

Transversal code-switching~\cite{heussen2025efficient}, in which transversal $\CNOT$\ gates facilitate teleportation of logical states between these codes, makes the EFTQC architecture that combines both of these codes universal.
(Other methods for realizing logical $H$ gates in the \cubecode\ are also possible, \eg, the methods described in Ref.~\cite{BerlinPaper}.)
These teleportation operations are the foundational primitive and one of the performance-limiting operations in our architecture.
Fault-tolerant circuits for teleporting between the codes are depicted in Figs.~\ref{fig:X_teleport_circ} and~\ref{fig:Z_teleport_circ}.  
The teleportations depend on the ability to fault-tolerantly prepare and measure encoded qubits in the Pauli $X$ and $Z$ bases, as described in the next section.


\subsection{State preparation and measurement}
\label{sec:spam}

The preparations and projective measurements we use for the RCA algorithm are preparations of logical $|000\>$ and $|{+}00\>$ in the \cubecode\ code, preparations of logical $|0{+}\>$ in the \squarecode\ code, projective measurements of $M_X$ in the \cubecode\ code, and projective measurements of $M_Z$ in the \squarecode\ code, each for arbitrary logical qubits in these code blocks.  
Of course, we also use destructive measurements of $M_Z$ to do final readouts of logical qubits, which can be done transversally for either code.

The canonical way to fault-tolerantly prepare the logical $|000\>$ state in the \cubecode\ code, given that it is a CSS code~\cite{Calderbank_1996, steane1996}, is to prepare $|0\>^{\otimes n}$ on the physical qubits and then fault-tolerantly measure the code's $X$ checks.  
For the \cubecode\ code, there is just one $X$ check, which can be fault-tolerantly measured by measuring the stabilizer generator $S_5$ in Table~\ref{tab:codes} twice, followed by a simple ``discard if unequal'' classical decoding algorithm.

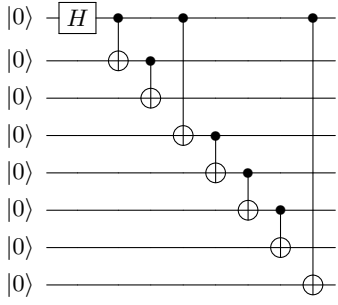
\begin{figure}[!ht]
    \centerline{
        \Qcircuit @C=0.5em @R=.7em {
        & \lstick{\ket{0}} & \gate{H} & \ctrl{1} & \qw & \ctrl{3} & \qw & \qw & \qw & \ctrl{7} & \qw \\
        & \lstick{\ket{0}} & \qw & \targ & \ctrl{1} & \qw & \qw & \qw & \qw & \qw & \qw  \\
        & \lstick{\ket{0}} & \qw & \qw & \targ & \qw & \qw & \qw & \qw & \qw & \qw  \\
        & \lstick{\ket{0}} & \qw & \qw & \qw & \targ & \ctrl{1} & \qw & \qw & \qw & \qw \\
        & \lstick{\ket{0}} & \qw & \qw & \qw & \qw & \targ & \ctrl{1} & \qw & \qw & \qw  \\
        & \lstick{\ket{0}} & \qw & \qw & \qw & \qw & \qw & \targ & \ctrl{1} & \qw & \qw \\
        & \lstick{\ket{0}} & \qw & \qw & \qw & \qw & \qw & \qw & \targ & \qw & \qw  \\
        & \lstick{\ket{0}} & \qw & \qw & \qw & \qw & \qw & \qw & \qw & \targ & \qw \\
        } 
    }
    \caption{Logical $\ket{{000}}$ state preparation for the \cubecode\ code. 
    Single $Z$ errors propagate only to checks, and no single $X$ error can grow higher than weight three, modulo the $X$ check.  
    Because the \(X\)-distance is four, the $Z$ checks detect all $X$ errors of weight three or less.
    Therefore any single $X$ error remains detectable despite potentially growing in weight.
    Thus this circuit preserves the overall distance of the code, albeit at the cost of rendering the $X$-distance of the circuit to be two.
    }
    \label{fig:prep000}
\end{figure}

An alternative fault-tolerant preparation procedure that uses fewer physical gates exploits the fact that, at the physical-qubit level, this codeword is just an eight-qubit cat state.  
Various fault-tolerant coherent state-preparation circuits for this exist. 
For example, the circuit in Fig.~\ref{fig:prep000} achieves this, as explained in the caption.

\begin{figure}[!h]
    \centerline{
        \Qcircuit @C=0.5em @R=.7em {
        & \lstick{\ket{0}} & \targ & \qw & \targ & \measure{M_Z} & \push{\ket{0}} & & \targ & \qw & \targ & \measure{M_Z} \\
        & \lstick{\ket{+}} & \ctrl{-1} & \ctrl{1} & \ctrl{-1} & \measure{M_X} & \ket{+} & & \ctrl{-1} & \ctrl{5} & \ctrl{-1} & \measure{M_X} \\
        & \lstick{\ket{0}} & \qw & \targ \qwx[1] & \qw & \qw & \qw & \qw & \qw & \qw & \qw & \qw  \\
        & \lstick{\ket{0}} & \qw & \targ \qwx[1] & \qw & \qw & \qw & \qw & \qw & \qw & \qw & \qw  \\
        & \lstick{\ket{0}} & \qw & \targ \qwx[1] & \qw & \qw & \qw & \qw & \qw & \qw & \qw & \qw  \\
        & \lstick{\ket{0}} & \qw & \targ & \qw & \qw & \qw & \qw & \qw & \qw & \qw & \qw \\
        & \lstick{\ket{0}} & \qw & \qw & \qw & \qw & \qw & \qw & \qw & \targ \qwx[1] & \qw & \qw  \\
        & \lstick{\ket{0}} & \qw & \qw & \qw & \qw & \qw & \qw & \qw & \targ \qwx[1] & \qw & \qw   \\
        & \lstick{\ket{0}} & \qw & \qw & \qw & \qw & \qw & \qw & \qw & \targ \qwx[1] & \qw & \qw  \\
        & \lstick{\ket{0}} & \qw & \qw & \qw & \qw & \qw & \qw & \qw & \targ & \qw & \qw  \\
        } 
    }
    \caption{Logical $\ket{{+00}}$ state preparation for the \cubecode\ code. 
    All physical qubits begin in the $\ket{0}$ state, then $\overline{X}_1$ and $S_5\overline{X}_1$ are measured using flag fault-tolerant measurements.
    This circuit is applied twice; in the first application, the $M_X$ measurements return random independent values.
    However, unless an error occurs, the second round of measurements should return the same values.
    An error is also detected if, at any point, an $M_Z$ measurement returns ${-}1$.
    }
    \label{fig:prep+00}
\end{figure}
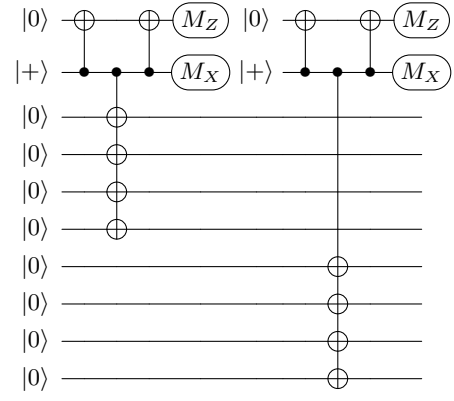

To prepare the logical $|{+}00\>$ state in the \cubecode\ code, we use the canonical approach for CSS codes, but instead of measuring $S_5$ twice, we measure $\overline{X}_1$ and $S_5\overline{X}_1$ twice each and apply the appropriate logical Pauli frame corrections to fix $+1$ values to both $S_5$ and $\overline{X}_1$.
Furthermore, to check the physical ancilla qubit used to realize these measurements for $X$-errors that could spread to the data qubits, we add an additional ``flag'' ancilla~\cite{PRXQuantum.1.010302} to each measurement.
The resulting circuit (which, as noted, must be run twice to be fault-tolerant to measurement errors) is shown in Fig.~\ref{fig:prep+00}.
The routine passes post-selection if all four $M_Z$ measurements return ${+}1$, and the pair of $M_X$ measurement outcomes returned the first time match the pair of $M_X$ measurement outcomes returned the second time.

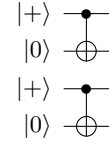
\begin{figure}[!h]
    \centerline{
        \Qcircuit @C=0.5em @R=1em {
        & \lstick{\ket{+}} & \ctrl{1} & \qw  \\
        & \lstick{\ket{0}} & \targ & \qw \\
        & \lstick{\ket{+}} & \ctrl{1} & \qw  \\
        & \lstick{\ket{0}} & \targ & \qw  \\
        } 
    }
    \caption{Fault-tolerant $\ket{{0+}}$ state preparation in the \squarecode\ code, as described in Ref.~\cite{gottesman2016quantumfaulttolerancesmall}.
    }
    \label{fig:prep0+}
\end{figure}

To prepare a logical $|0{+}\>$ state in the \squarecode\ code, we again note that the physical-qubit state that corresponds to this is simple.  
In this case, as also noted in Ref.~\cite{gottesman2016quantumfaulttolerancesmall}, the state is a pair of disjoint physical Bell states. 
The circuit shown in Fig.~\ref{fig:prep0+} is a fault-tolerant preparation circuit for this state because the $\CNOT$ gates can only spread Pauli errors to which the initial Pauli eigenstates are immune, ensuring that any single-error remains detectable.

To measure $M_X$ on a target logical qubit in the \cubecode\ code, one can use a single application of the circuit in Fig.~\ref{fig:prep+00}, adjusted to measure $\overline{X}_1$, $\overline{X}_2$, or $\overline{X}_3$ as desired, with the existing state of the \cubecode\ code used in place of the $|0\>^{\otimes 8}$ depicted in the circuit.  
A single application of this circuit suffices because the $S_5$ stabilizer of the \cubecode\ code is inferred from the parity of the ancilla $M_X$ measurements and an error is detected if either flag qubit measurements return ${-}1$.

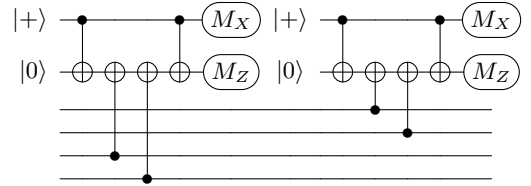
\begin{figure}[!h]
    \centerline{
        \Qcircuit @C=0.5em @R=0.7em {
       & \lstick{\ket{+}} & \ctrl{1} & \qw & \qw & \ctrl{1} & \measure{M_X} & \push{\ket{+}} & & \ctrl{1} & \qw & \qw & \ctrl{1} & \measure{M_X} \\
       & \lstick{\ket{0}} & \targ & \targ & \targ & \targ & \measure{M_Z} & \push{\ket{0}} & & \targ & \targ & \targ & \targ & \measure{M_Z}\\
        &  & \qw & \qw & \qw & \qw & \qw & \qw & \qw & \qw &  \ctrl{-1} & \qw & \qw & \qw \\
        &  & \qw & \qw & \qw & \qw & \qw & \qw & \qw & \qw & \qw & \ctrl{-2} & \qw & \qw \\
        &  & \qw & \ctrl{-3} & \qw & \qw & \qw & \qw & \qw & \qw & \qw & \qw & \qw & \qw \\
        &  & \qw & \qw & \ctrl{-4} & \qw & \qw & \qw & \qw & \qw & \qw & \qw & \qw & \qw \\
        } 
    }
    \caption{Fault-tolerant measurement of the $\overline{Z}_2$ operator in the \squarecode\ code.
    The two $M_Z$ measurements return the values of $\overline{Z}_2$ and $S_2\overline{Z}_2$, such that $S_2$ is obtained from the parity of these measurements.
    The $M_X$ measurements serve as flags and an error is detected if either returns ${-}1$.}
    \label{fig:Mz_422}
\end{figure}

To measure $M_Z$ on a target logical qubit $j$ in the \squarecode\ code, one can use a similar procedure, say, by measuring $\overline{Z}_j$ and $S_2\overline{Z}_j$, such that $S_2$ is obtained from the parity of these measurements and a flag qubit catches extra errors.  
A circuit for doing this for $j=2$ is depicted in Fig.~\ref{fig:Mz_422}.

\section{Compiling the RCA algorithm to our architecture}
\label{sec:RCA}

We first briefly review the quantum RCA algorithm of Cuccaro \etal~\cite{cuccaro2004newquantumripplecarryaddition} and then describe how to compile this into our architecture.
Many other quantum adder algorithms have been developed since the Cuccaro adder, such as those referenced in Ref.~\cite{Wang:2025a}; deploying them and other algorithms beyond addition on our EFTQC architecture could also be interesting.
We then discuss how the initial classical carry bit used in the RCA algorithm can be used as an additional ``algorithmic check'' to make its implementation more robust.

\subsection{Ripple-Carry Addition}

\begin{figure}[!h]
    \centering
    \includegraphics{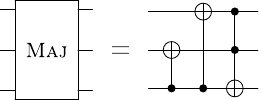}
    \caption{The \textsc{Maj} circuit.}
    \label{fig:maj}
\end{figure}

\begin{figure}[!h]
    \centering
    \includegraphics{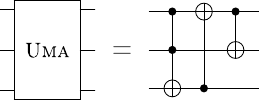}
    \caption{The \textsc{Uma} circuit.}
    \label{fig:uma}
\end{figure}

The quantum RCA algorithm of Cuccaro \etal~\cite{cuccaro2004newquantumripplecarryaddition} is just a classical RCA algorithm that uses the reversible $\CNOT$\ and $\TOF$ gates, where the $\TOF$\ gate is the Toffoli gate, a.k.a.\ controlled-controlled-$X$, or \textit{CCX}. 
This RCA algorithm is built from two subroutines, \textsc{Majority} (\textsc{Maj}) and \textsc{UnMajority and Add} (\textsc{Uma}), each consisting of a single Toffoli gate and two $\CNOT$ gates. 
These subroutines are shown in Figs.~\ref{fig:maj} and~\ref{fig:uma}.

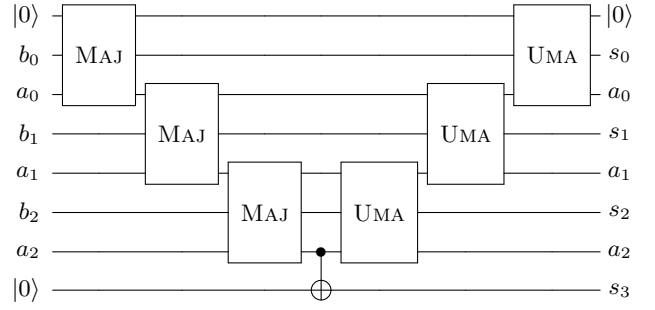
\begin{figure}[!h]
    \centerline{
\Qcircuit @C=0.4em @R=0.7em {
        \lstick{\ket{0}} & \multigate{2}{\textsc{Maj}} & \qw & \qw & \qw & \qw & \qw & \multigate{2}{\textsc{Uma}} & \qw & & \ket{0} \\
        \lstick{b_0} & \ghost{\textsc{Maj}} & \qw  & \qw & \qw & \qw & \qw & \ghost{\textsc{Uma}} & \qw & & s_0 \\
        \lstick{a_0} & \ghost{\textsc{Maj}} & \multigate{2}{\textsc{Maj}}  & \qw & \qw & \qw & \multigate{2}{\textsc{Uma}} & \ghost{\textsc{Uma}} & \qw & & a_0
        \\
        \lstick{b_1} & \qw & \ghost{\textsc{Maj}} & \qw & \qw & \qw & \ghost{\textsc{Uma}} & \qw & \qw & & s_1
         \\ 
        \lstick{a_1} & \qw & \ghost{\textsc{Maj}} &  \multigate{2}{\textsc{Maj}} & \qw & \multigate{2}{\textsc{Uma}} & \ghost{\textsc{Uma}}  & \qw & \qw & & a_1
         \\ 
        \lstick{b_2} & \qw & \qw & \ghost{\textsc{Maj}} & \qw & \ghost{\textsc{Uma}} & \qw  & \qw & \qw & & s_2
        \\
        \lstick{a_2} & \qw & \qw & \ghost{\textsc{Maj}} & \ctrl{1} & \ghost{\textsc{Uma}} & \qw  & \qw & \qw & & a_2
        \\
        \lstick{\ket{0}} & \qw & \qw & \qw & \targ & \qw & \qw & \qw & \qw  & & s_3 
    }
    }
    \caption{Example $3$-bit RCA circuit computing $a+b=s$ constructed from descending $\textsc{Maj}$s followed by ascending $\textsc{Uma}$s. 
    In general, $m$-bit RCA uses $m$ $\textsc{Maj}$ and $\textsc{Uma}$ gates.
    At the end of the circuit, the topmost workspace qubit is uncomputed back to $\ket{0}$.
    Testing whether this reset happens is an algorithmic check, independent of any QEC code that may be used to realize the circuit on logical qubits.}
    \label{fig:rca3}    
\end{figure}

An example $3$-bit RCA circuit is shown in Fig.~\ref{fig:rca3}, from which the general algorithm can easily be deduced.

The general algorithm can be understood for classical $m$-bit addends $a$ and $b$ as follows.  
Starting from bit $i=0$ and advancing to bit $i=m-2$, $\textsc{Maj}$ takes the $i$-th bit from each bitstring, $a_i$ and $b_i$, and the $i$-th carry bit $c_i$ and computes $c_i \oplus a_i$ into the bit where $c_i$ was stored, $s_i = a_i \oplus b_i$ into the bit where $b_i$ was stored, and the next carry bit $c_{i+1}$ into the bit where $a_i$ was stored.
The full addition circuit is then performed by applying $\textsc{Maj}$ to all the bits in $a$ and $b$ sequentially, using the outputs $c_i$ as inputs to next $\textsc{Maj}$.
The final carry bit is copied onto an ancilla with the single $\CNOT$, then the $\textsc{Maj}$ are uncomputed with the $\textsc{Uma}$ circuits.  
The $\textsc{Uma}$ circuit isn't just the inverse of the $\textsc{Maj}$ circuit; if it were, this algorithm would just restore the input to its original state, recording only the final carry bit $s_{m} = c_m$.  
Instead, it propagates the carry bits backwards, restoring the $a_i$ bits back to their original locations while preserving the $s_i$ bits, until it finally restores the initial carry bit $c_0$ to $0$.
The circuit acts the same way termwise in superposition on addends in quantum states.

The $m$-bit RCA algorithm uses two additional qubits beyond the \(2m\) needed to store the bitstrings \(a\) and \(b\).
The first is the initial qubit on the top line that represents the initial carry bit $c_0$, which must be prepared in the classical basis state $|0\>$, regardless of whether the addends are in a nontrivial quantum state or not.  
The second is the final carry bit $c_m$ on the bottom line, which also should be prepared in the state $|0\>$.
Generally, the final carry bit will end up in an entangled state with the various superposition terms in the addends, but the initial carry bit $c_0$ will always be restored to the product state $|0\>$.  
This provides an additional check that we call an ``algorithmic'' check, as we are able to test whether $c_0$ has been reset to $|0\>$.

\subsection{Compilation}

\begin{figure}[!h]
    \centering
    \includegraphics{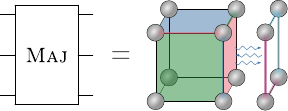}
    \caption{The \textsc{Maj} operation compiled into our architecture consists of three teleportation operations between an \cubecode\ and a \squarecode\ code.}
    \label{fig:maj_sia}
\end{figure}

The Cuccaro \etal\ $m$-bit RCA algorithm has a particularly natural compilation into our architecture.  We do not claim that it is optimal, but it dovetails well with the structure of the codes that our architecture uses.
To begin, we initialize the logical qubits in one \cubecode\ code into $\ket{{0a_0b_0}}$ and the logical qubits in $m-1$ \cubecode\ codes into $\ket{{+a_ib_i}}$.  On the first block, the first two $\CNOT$ gates in the \textsc{Maj} subroutine are implemented by passive permutations.  
To perform the succeeding $\TOF$ gate, we teleport the target qubit to a \squarecode\ code for a transversal Hadamard gate, teleport it back to the \cubecode\ code for a transversal \CCZ\ gate, then teleport it back to the \squarecode\ code for a second transversal Hadamard gate.
This is depicted schematically in Fig.~\ref{fig:maj_sia}.

\begin{figure}[!h]
    \centering
    \includegraphics{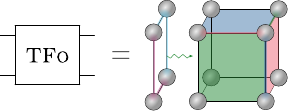}
    \caption{The teleport forward (\textsc{TFo}) operation consists of teleporting a state from a \squarecode\ code onto a \cubecode\ code.}
    \label{fig:tfo}
\end{figure}

At this point, the \textsc{Maj} is completed, but the bottom qubit is in a \squarecode\ code, not an \cubecode\ code.
This qubit stores the carry bit needed as an input for the next \textsc{Maj} operation.
Instead of teleporting this state back onto the \cubecode\ code from which it originated, we teleport it forward onto first logical qubit of the next \cubecode\ block that stores $|{+}a_1b_1\>$.  
Fig.~\ref{fig:tfo} depicts this ``teleport forward'' operation that we denote by \textsc{TFo}, again in a schematic fashion.

We continue this descending cascade of operations until the final carry-bit is in a \squarecode\ code.  
We then bring in a second \squarecode\ code block whose logical qubits are initialized to $\ket{{0+}}$, and apply a transversal $\CNOT$ between the two \squarecode\ blocks, copying this carry-bit $s_m$ into the first logical qubit of the second block.
We are now ready to begin the ascending cascade of \textsc{Uma} operations.

\begin{figure}[!h]
    \centering
    \includegraphics{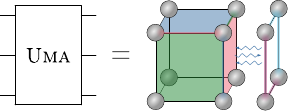}
    \caption{The \textsc{Uma} operation compiled into our architecture consists of three teleportation operations in between an \cubecode\ and a \squarecode\ code.}
    \label{fig:uma_sia}
\end{figure}

We implement the Toffoli gates in the \textsc{Uma} operations in a similar fashion to the \textsc{Maj} gates. 
However, we begin each \textsc{Uma} gate in a \squarecode\ code and end in an \cubecode\ code.
Hence, we implement the Toffoli gate by first applying a transversal Hadamard gate, teleporting to the \cubecode\ code for the transversal \CCZ\ gate, teleporting back to the \squarecode\ code for a second transversal Hadamard gate, teleporting back to the  \cubecode\ code, and finally performing the remaining $\CNOT$s with passive physical qubit relabelings.  
We depict this schematically in Fig.~\ref{fig:uma_sia}.

\begin{figure}[!h]
    \centering
    \includegraphics{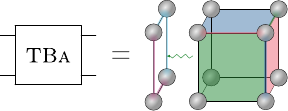}
    \caption{The teleport backwards (\textsc{TBa}) operation consists of teleporting a state from an \cubecode\ code onto a \squarecode\ code.}
    \label{fig:tba}
\end{figure}

Similarly to how we propagated the relevant logical qubits forward to the next \cubecode\ code block using $\textsc{TFo}$ with $\textsc{Maj}$, we propagate the relevant logical qubits backwards to the previous \cubecode\ code block using a ``teleport backwards'' operation that we denote by \textsc{TBa}.  
We depict this schematically in Fig.~\ref{fig:tba}.

Cascading these ascending steps together completes the algorithm.  
We can perform a logical $Z$ measurement on $c_0$ to see that it is in the state $0$ as an algorithmic check, if we desire.  
If no further operations are to be done on the outputs, we can also measure the logical qubits in the $Z$ basis destructively at this point.  
As usual, a destructive $Z$ measurement also allows us to infer the final $Z$ stabilizer measurements in a measurement-error-free way, as they are computed classically from the individual physical-bit outcomes.

\begin{figure*}[!t]
    \centering
    \includegraphics[width=0.95\linewidth]{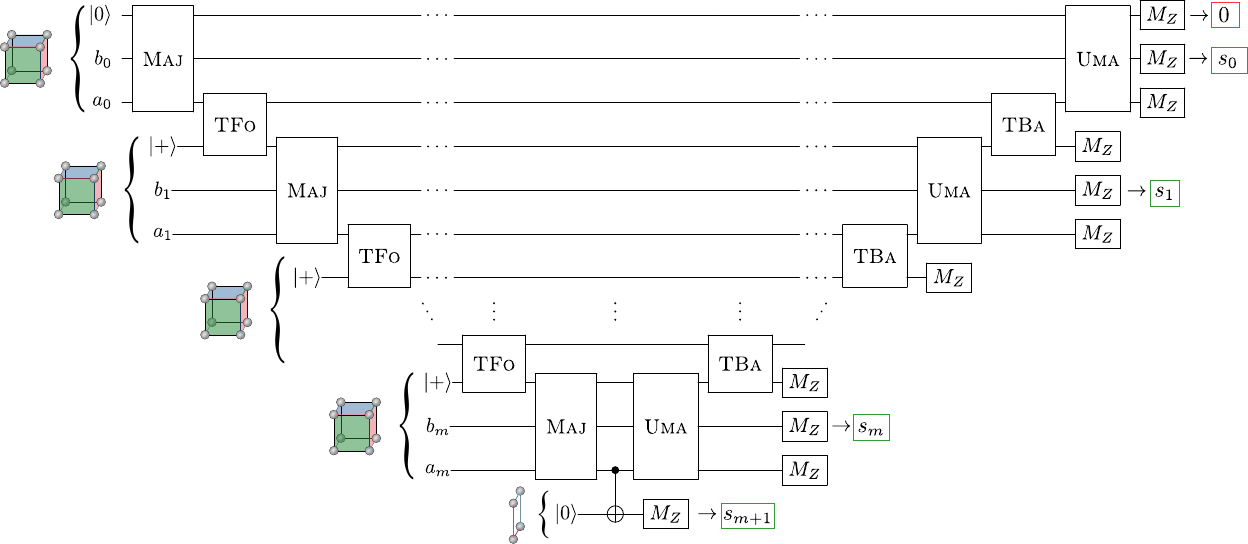}
    \caption{Compilation of $m$-bit RCA for classical bit-strings, consisting of cascading \textsc{Maj} and \textsc{TFo} operations, followed by ascending \textsc{TBa} and \textsc{Uma} operations. 
    The individual bits in the output bitstring $s$ are circled in green and the algorithmic check is circled in red.
    Note that while this circuit displays certain operations in parallel, every operation is serialized when executed on Helios-1.}
    \label{fig:rca_circuits}
\end{figure*}

To reduce the exposure of our compiled RCA algorithm to noise, we make three modifications to the general compilation algorithm we have described so far.  
First, because we only consider classical addend inputs in our experiments, we prepare them on-demand, right before they are fed into \textsc{Maj} gates.  
Second, because we are not continuing after the $m$-bit addition, we measure each bit immediately after the relevant \textsc{Uma} gate is executed.  
This leads to the ``V''-shaped circuit depicted in Fig.~\ref{fig:rca_circuits} and allows us to reuse the qubits allocated to the final \squarecode\ code storing $s_m$ for the subsequent teleportation operations.  
In fact, as we will see later, we found that an implementation that uses the same four physical qubits for all teleportation operations, each of which is only needed for a short-lived interaction, was the best choice on Quantinuum's Helios-1 hardware.

Most significantly, we do not perform syndrome extraction after each of the operations in the list above as a generic fault-tolerant protocol would.  
Instead, we only use the syndrome bits generated during the projective logical measurements, each of which only reveals some, but not all, of the syndrome, namely $S_5$ for the \cubecode\ code and $S_2$ for the \squarecode\ code.  
Because all logical gates are implemented transversally, the only places where errors can spread are in the preparation and teleportation subroutines that use these projective measurements, where the fault-tolerant implementation of these ensures that the code distance is still two. 
However, the effective physical error rate in logical qubits might be much higher than if they were checked more regularly.
This is particularly noticeable in the top logical qubit storing $c_0$, which undergoes $\bigO(m)$ logical identity gates between partial syndrome extractions in the \textsc{Maj} and \textsc{Uma} subroutines.  
We expect this logical qubit to have an effective physical error rate that is $\bigO(m)$ times larger than that of the final logical qubit that holds the last carry bit $s_m$.

\subsection{Co-design with the algorithmic check}

While the overall fault distance is two in our architecture, meaning that single-qubit errors can be detected, in many circuit locations the effective fault-distance is four, meaning that three-qubit errors can be detected.
This is because the terminating measurements are all in the $Z$-basis, implying only logical-$X$ errors flip the outcomes of these circuits.
While the \cubecode\ has a code distance of two, it has an $X$-distance of four.
Since there are logical Hadamard gates in the circuit, a logical-$Z$ error could turn into a logical-$X$ error.
This happens if a logical-$Z$ error occurs in between a pair of logical Hadamard gates conjugating a logical $\CCZ$ gate into a Toffoli gate.
However, any single logical-$Z$ error occurring in one of these locations will ultimately propagate all the way to the algorithmic check, which can detect it.  
Thus, at least two logical-$Z$ errors must occur to pass undetected by the algorithmic check.

\begin{figure}[!h]
    \centering
    \includegraphics[width=0.9\linewidth]{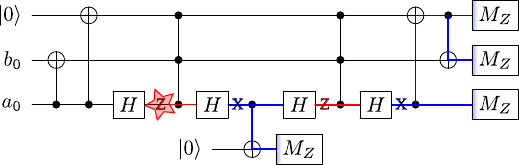}
    \caption{Example of the algorithmic check in single-bit RCA with the Toffoli gates decomposed as $H$--$\CCZ$--$H$. A logical $Z$ error occurring in the middle of this sequence gets conjugated into an $X$ error, which flips the outcome of the terminating $Z$-basis measurements and causes the adder to fail. However, these same errors also propagate to the top ancilla qubit, flipping the outcome of the algorithmic check. For arbitrary $m$-bit addition, the \textsc{Uma} operations similarly propagate these errors to the algorithmic-check qubit. }
    \label{fig:AC_example}
\end{figure}

We show an example of this in Fig.~\ref{fig:AC_example} for a $1$-bit RCA circuit with the Toffoli gates decomposed into \CCZ\ and Hadamard gates.
In this example, a logical $Z$-error occurring in the middle of one of these gate sequences gets transformed into a logical $X$-error by the following Hadamard gate, propagating to multiple output bits.
However, since this error also flips the top most ancilla qubit, it is detected by the algorithmic check.
When generalized to the $m$-bit RCA algorithm, the cascading/ascending structure of the \textsc{Maj}/\textsc{Uma} oracles ensures any single logical $Z$-error occurring in one of these locations propagates up to the algorithmic check qubit.

Because the $X$-distance of the \squarecode\ code is two, it is not surprising that there are numerous circuit locations where two physical-$X$ errors will cause the circuit to fail.
This happens, for example, whenever encoded information is stored in the \squarecode\ code and transversal Hadamard gates are applied.
As such, we attempt to minimize the time spent with states stored in a \squarecode\ code.

\section{Results}
\label{sec:results}

\subsection{Proxy Circuit Simulation}

In order to guide specific implementation choices for experimental runs on Helios-1, we constructed proxy Clifford circuits that could be simulated efficiently using Helios-1E, Quantinuum's emulator containing their noise model of Helios-1.  
To do this, we replaced each transversal $\CCZ$ gate with two transversal logical $CZ$ gates and coded up the entire algorithm in Quantinuum's \textsc{guppy} language~\cite{koch2025guppypythonicquantumclassicalprogramming}.
We simulated many different equivalent proxy circuits before settling on the specific implementation with the best performance.

In our final implementation choice, we reused the same four qubits for all  \squarecode\ code blocks, resetting the logical $\ket{{+}}$ qubit in the \squarecode\ in the relevant teleportation circuits (Figs.~\ref{fig:X_teleport_circ} and~\ref{fig:Z_teleport_circ}) in the middle of the \textsc{Maj} and \textsc{Uma} operations with a projective $\overline{X}$ measurement.
This requires only a single logical $\ket{{0+}}$ preparation at the beginning of the circuit, but $2m$ $\overline{X}$ measurements.

Throughout the circuit, we use two physical ancilla qubits to realize each logical projective measurement, though they need not be the same physical qubits each time.
Thus at least $8m + 10$ physical qubits were used for $m$-bit RCA circuits, with the exact number set by the compiler for Helios-1.

In addition to using the proxy-circuit simulations to finalize our implementation choices, we used them to guide the design of the \textsc{guppy} programs.
Quantinuum's proprietary compiler compiles each \textsc{guppy} program into the native Helios-1 gate set and appears sensitive to the specific way the program is expressed.
Therefore, even after we finalized the implementation details, we simulated several functionally equivalent \textsc{guppy} programs and selected the design that achieved the best performance on the proxy circuits.

\subsection{Resource Counts}

\begin{figure}[!h]
    \centering
    \includegraphics[width= \linewidth]{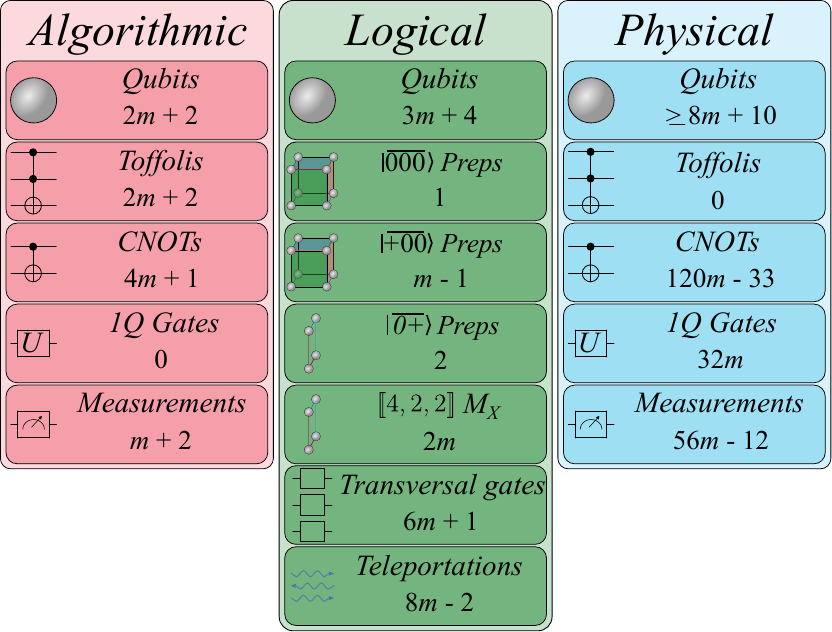}
    \caption{Counting the required resources for \(m\)-bit RCA at three different levels: algorithm, logical and physical.
    The physical counts represent the size of the \textsc{guppy} programs submitted to Helios-1.
    The qubit and gate counts after compiling these programs into Helios-1's native gate set are not shown.
    }
    \label{fig:counts}
\end{figure}

Here we report the size of the compiled RCA circuits submitted to Helios-1. 
We evaluate circuit size across three different cost models: algorithmic, logical, and physical. The algorithmic cost model counts the qubits and operations specified by the algorithm prior to compilation, using the standard gate set as depicted in Fig.~\ref{fig:rca3}. 
The logical cost model tracks the logical qubits and architectural primitives required to implement the algorithm on our EFTQC architecture.
The relevant primitives here are state preparations, teleportation operations, and transversal gates. 
Finally, the physical cost model counts the physical qubits and gates required to realize these architectural primitives, again in the gate set consisting of Clifford preparations, gates, and measurements along with Toffoli gates. 
These physical counts correspond to the size of the \textsc{guppy} programs submitted to Helios-1.
Counts at all three levels are reported in Fig.~\ref{fig:counts}. 
We note that an additional compilation stage maps these programs to Helios-1's native gate set, so our use of the word ``physical'' is somewhat of a misnomer.
This stage is performed by Helios-1's compiler and determines the final number of qubits and gates used on the device.  
We did not have access to the final compiled circuits at the truly physical-qubit level, so we do not report the resource costs at that level.

\subsection{Logical RCA Results}

\begin{figure*}[!ht]
    \centering
    \includegraphics[width=2.0\columnwidth]{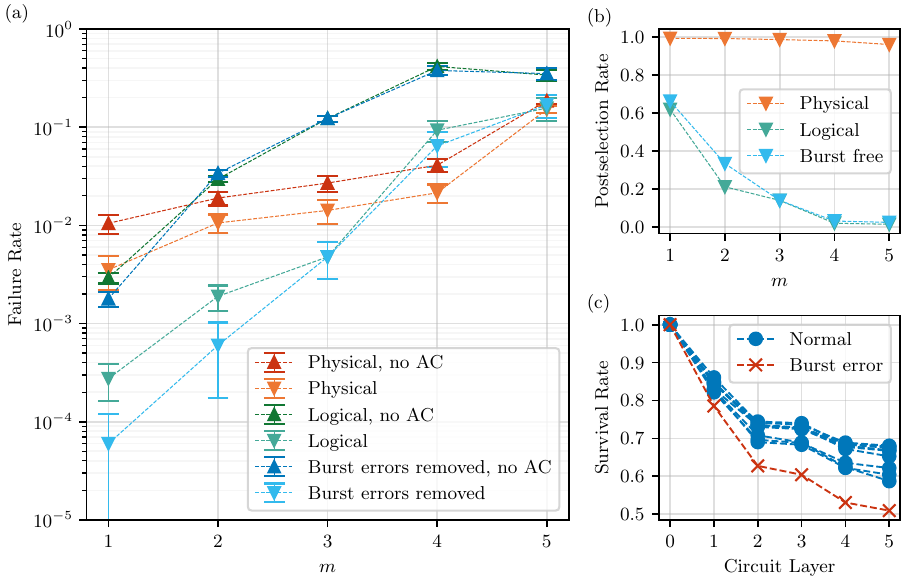}
    \caption{Results for RCA execution on Helios-1. 
    In (a) we compare physical and logical error rates, both with and without the algorithmic check (AC). We denote the logical data sets which include all post-selected data as ``Logical'' and the data with burst-errors removed as ``Burst errors removed.''
    When including the AC, the logical failure rates are below physical failure rates for $m = 1$ to $3$.
    The error bars reported here show the binomial standard error for each data set.
    In (b) we show the post-selection rates combining the error detection and AC. 
    We do not include the shots discarded due to burst-errors in this plot. 
    In (c) we show an example of a job containing burst errors. 
    Each data set consists of $\sim\!10^3-10^4$ shots. We track the average survival rate of shots through an RCA-circuit at fixed $m$. 
    Here one job is identified as containing burst errors since its survival rate decays much faster the other jobs.}
    \label{fig:rca_data}
\end{figure*}

We submitted our circuits to Helios-1 in jobs containing batches of at least $10^3$ shots, but not more than $10^4$ shots, continuing to submit jobs until at least three failures were detected.
For each job, we selected the classical addends $a$ and $b$ independently and uniformly at random from among $m$-bit strings.  
However, for each value of \(m\), we ensured that the maximum values of the \(m\)-bit addends (\viz, $2^m-1$) were selected at least once.  
These choices corresponded to logical-qubit Pauli frame choices; in practice, we ran every circuit on the $|0\>^{\otimes m}$ input for each addend and propagated the Pauli frame classically through the circuit in the analysis.
Each shot within a job then ran the same \(m\)-bit RCA circuit with the same addends.
When all syndrome measurements detected no errors, we declared a shot fully post-selected. 
We declared a fully post-selected shot a success if the final outcome reported \(s = a + b\), and a failure otherwise. 
A failure which passed all syndrome measurements but failed the final algorithmic check was declared a failure without the AC, but post-selected against when including the AC.
In principle, we could have also checked that each bit of $a$ going in matched each bit of $a$ going out as additional ``algorithmic checks,'' because each $a$ is a classical input.  
However, we did not do this as it would unfairly make the algorithm look better than it would have been---had we used quantum addend inputs instead, which could have been in a superposition or entangled state with other qubits, the RCA outputs would have been unable to be  verified that way.

Whenever we detected an error mid-circuit, the circuit was terminated and the circuit layer where the error was detected was returned.
This allowed us to track average shot survival rates vs.\ circuit layer. 
By examining the average survival rates of individual job submissions, we identified some apparent burst-like errors. 
For example, at fixed $m$, jobs implementing the same $m$-bit adder typically exhibited comparable average survival rates. 
However, in some jobs, the survival rate decayed much more rapidly, suggesting that Helios-1 was not operating at its typical performance during those runs. 
We therefore report failure rates both using all data that survived post-selection and after discarding all shots from jobs flagged as exhibiting burst errors, regardless of whether those shots survived post-selection.

Our results are summarized in Fig.~\ref{fig:rca_data}. 
We report the logical failure rates and standard error both with and without the algorithmic check, and both with and without removing burst-error events.
We compare these rates against the failure rates of the corresponding physical implementations, again with and without the algorithmic check. 
Without the algorithmic check, the logical circuits perform below pseudo-threshold only for single-bit addition. 
With the algorithmic check, however, the logical failure rates remain below the physical failure rates through $m=3$. 
While the algorithmic check does improve the physical failure rates, it has a much larger impact on the logical circuits, likely reflecting the synergy it has with the error-detecting codes as described earlier.
We also report post-selection rates both with and without removing burst-error events. 
In both cases, the post-selection rate starts slightly above $60\%$ for $m=1$ and decays exponentially to approximately $1\%$ at $m=5$.  
Thus, although Helios-1 is capable of holding much larger instances than $m=5$ with its 98 qubits, we found ourselves limited by the yield rate afforded by the underlying fidelity of the operations and the low distance of the codes used in our EFTQC architecture.  
By contrast, post-selection rates for the physical circuits using the algorithmic check remain above $95\%$ throughout.
As noted earlier, for each value of \(m\), we continued submitting jobs until we observed at least three failures. 
However, we identified the burst-error events only after data collection had concluded. 
As a result, after removing these events, the data for \(m=1\) and \(m=2\) contain fewer than three failures: one and two failures, respectively. 
The corresponding error bars in Fig.~\ref{fig:rca_data} reflect these small failure counts.

The $m = 5$ bit addition data, for both the logical and physical failure rates, manifestly departs from the trends observed for $m = 1$ to $4$.
We repeated these circuits multiple times, on different days and with varying input bit strings, and consistently observed the same behavior.  
We remark on this unusual behavior in the next section.

\section{Summary and Conclusion}
\label{sec:conclusion}

We compiled Cuccaro's quantum ripple-carry adder, for instances of up to five-bit addends, onto an early fault-tolerant quantum computing architecture that we developed and deployed on Quantinuum's trapped-ion platform Helios-1. 
Using the algorithmic check provided by this algorithm, we showed that the fault-distances of many vulnerable circuit locations are increased to four, well beyond the distance of two of the QEC codes that the EFTQC architecture utilizes. 
By tracking syndrome data in batch jobs, we were able to identify runs subject to atypical burst errors and compare logical failure rates both with and without removing such runs.
On runs that passed the algorithmic check and in which no errors were detected, we observed logical failure rates improving upon physical failure rates for $m \leq 3$.
At $m = 1$ our burst-error-filtered data points to a logical failure rate of $(6.0 \pm 5.9) \times 10^{-5}$, which is a factor more than 50 below the corresponding physical failure rate of $(3.5 \pm 1.3) \times 10^{-3}$.  

At $m=4$, we found that our encoded algorithm performed worse than the physical one, but only slightly.  
We suspect that this is because we did not run active syndrome extraction and post-selection circuits on idling logical qubits in our implementation.  
Thus, there were $\bigO(m^2)$ logical identities in the circuit whose syndromes were not examined.  
The impact of this was to increase the effective physical error rate.  
We did not perform these syndrome extraction steps for two reasons.  
First, doing so would reduce the yield of post-selection beyond its already meager level at $m=4$ (the yield drops exponentially in the number of post-selection steps).
More importantly, the hardware architecture of Helios-1 does not permit parallel syndrome extraction operations on the multiple code blocks we needed to keep alive during the algorithm.  
Had we tried to perform these syndrome extractions serially, it would lead to even more logical idle steps in a runaway cascade.  

The $m=5$ data we observed seems to defy the natural extrapolations one would make with our lower-$m$ data.
While we don't have a complete theory as to what could explain this, we suspect that a significant component of this is an extension of the effects we noted for the $m=4$ case.

We restricted our study to classical bit-string inputs, which allowed us to reduce idling time by initializing code blocks immediately before they were needed, rather than preparing superpositions over length-$m$ bit strings. 
If we had used quantum inputs, or embedded the RCA circuits as a subroutine within a larger quantum algorithm, we likely would have made different design choices. 
For example, with general quantum inputs, one would need to perform state preparation at the beginning of the circuit rather than initialize the \cubecode\ blocks one at a time in the middle of the circuit. 
In that context, we might have chosen to reduce idling time by using variants of the \textsc{Maj} and \textsc{Uma} subroutines that allow greater parallelism, as noted in the Cuccarro adder paper itself~\cite{cuccaro2004newquantumripplecarryaddition}. 
As another example, if we had executed our RCA circuits as subroutines within some larger algorithm, we might not have been able to measure the logical qubits as early as we did in our experiments. 
In that case, we might have chosen to store the final carry bit on a \cubecode\ code rather than a \squarecode\ code.

While Takahashi \etal\ seemingly optimized the Cuccaro adder by removing the initial carry bit in Ref.~\cite{Takahashi:2005a}, it overlooked the value that this ``gauge'' degree of freedom imputed to our EFTQC implementation of RCA addition.  
Given our observations, we believe that, as a general strategy for resource-efficient fault-tolerant quantum algorithms, one should co-design both the algorithm and the underlying architecture to achieve the best results.

\section*{Data Availability}

The data supporting the findings of this work is available from the corresponding author upon reasonable request.

\begin{acknowledgments}

In alphabetical order, we thank 
Riley Chien,
Shival Dasu,
Dave Hayes,
Cole Maurer,
Setso Metodi,
Mason Rhodes,
and 
Ciaran Ryan-Anderson
for helpful technical discussions.
JSN and ADB were supported by the DOE Office of Fusion Energy Sciences ``Foundations for quantum simulation of warm dense matter'' project.
Access to Quantinuum's hardware was made possible by the Advanced Simulation and Computing (ASC) program of the National Nuclear Security Administration.
Support is also acknowledged from the U.S. Department of Energy, Office of Science, National Quantum Information Science Research Centers, Quantum Systems Accelerator (Award No. DE-SCL0000121).

This article has been co-authored by employees of National Technology \& Engineering Solutions of Sandia, LLC under Contract No.\ DE-NA0003525 with the U.S. Department of Energy (DOE). The authors own all right, title and interest in and to the article and are solely responsible for its contents. The United States Government retains and the publisher, by accepting the article for publication, acknowledges that the United States Government retains a non-exclusive, paid-up, irrevocable, world-wide license to publish or reproduce the published form of this article or allow others to do so, for United States Government purposes. The DOE will provide public access to these results of federally sponsored research in accordance with the DOE Public Access Plan \url{https://www.energy.gov/downloads/doe-public-access-plan}.

\end{acknowledgments}

\bibliographystyle{landahl}

\bibliography{references,landahl}

@ARTICLE{Gottesman:1998a,
  author = {Daniel Gottesman},
  title = {Theory of fault-tolerant quantum computation},
  journal = {Phys. Rev. A},
  year = {1998},
  volume = {57},
  pages = {127--137},
  number = {1},
  month = {Jan.},
  comment = {e-print has slightly different title.},
  doi = {10.1103/PhysRevA.57.127},
  eprint = {quant-ph/9702029}
}

@Article{Preskill:2018b,
  author  = {John Preskill},
  journal = {Quantum},
  title   = {Quantum computing in the {NISQ} era and beyond},
  year    = {2018},
  pages   = {79},
  volume  = {2},
  doi     = {10.22331/q-2018-08-06-79},
  eprint  = {1801.00862},
}

@ARTICLE{Shi:2003a,
  author = {Yaoyun Shi},
  title = {Both {T}offoli and controlled-{NOT} need little help to do universal
	quantum computation},
  journal = {Quant. Inf. Comp.},
  year = {2003},
  volume = {3},
  pages = {84--92},
  number = {1},
  month = {Jan.},
  eprint = {quant-ph/0205115},
  url = {http://www.rintonpress.com/xqic3/qic-3-1/084-092.ps}
}

@Article{Self:2024a,
  author  = {Self, C.N. and Benedetti, M. and Amaro, D.},
  journal = {Nat. Phys.},
  title   = {Protecting expressive circuits with a quantum error detection code},
  year    = {2024},
  pages   = {219},
  volume  = {20},
  doi     = {10.1038/s41567-023-02282-2},
  eprint  = {2211.06703},
}

@article{Takahashi:2005a,
  author = {Y. Takahashi and N. Kunihiro},
  title = {A linear-size quantum circuit for addition with no ancillary qubits},
  journal = {Quantum Info. Comput.},
  volume = {5},
  pages = {440},
  year = {2005},
  url = {https://www.rintonpress.com/xqic5/qic-5-6/440-448.pdf},
}

@article{Wang:2025a,
    author = {Wang, Siyi and Li, Xiufan and Lee, Wei Jie Bryan and Deb, Suman and Lim, Eugene and Chattopadhyay, Anupam},
    title = {A comprehensive study of quantum arithmetic circuits},
    journal = {Philosophical Transactions of the Royal Society A: Mathematical, Physical and Engineering Sciences},
    volume = {383},
    number = {2288},
    pages = {20230392},
    year = {2025},
    month = {01},
    issn = {1364-503X},
    doi = {10.1098/rsta.2023.0392},
    url = {https://doi.org/10.1098/rsta.2023.0392},
    eprint = {2406.03867}
}

@misc{aharonov2003simple,
      title={A Simple Proof that Toffoli and Hadamard are Quantum Universal},
      author={Dorit Aharonov},
      year={2003},
      eprint={quant-ph/0301040},
      archivePrefix={arXiv},
      primaryClass={quant-ph}
}

@misc{reichardt2024demonstrationquantumcomputationerror,
      title={Demonstration of quantum computation and error correction with a tesseract code}, 
      author={Ben W. Reichardt and David Aasen and Rui Chao and Alex Chernoguzov and Wim van Dam and John P. Gaebler and Dan Gresh and Dominic Lucchetti and Michael Mills and Steven A. Moses and Brian Neyenhuis and Adam Paetznick and Andres Paz and Peter E. Siegfried and Marcus P. da Silva and Krysta M. Svore and Zhenghan Wang and Matt Zanner},
      year={2024},
      eprint={2409.04628},
      archivePrefix={arXiv},
      primaryClass={quant-ph}
}

@ELECTRONIC{Campbell:2016a,
  author = {Earl T. Campbell},
  month = {Sep. 26},
  year = {2016},
  title = {The smallest interesting colour code},
  howpublished = {Blog post},
  url = {https://earltcampbell.com/2016/09/26/the-smallest-interesting-colour-code/}
}

@misc{cuccaro2004newquantumripplecarryaddition,
      title={A new quantum ripple-carry addition circuit}, 
      author={Steven A. Cuccaro and Thomas G. Draper and Samuel A. Kutin and David Petrie Moulton},
      year={2004},
      eprint={quant-ph/0410184},
      archivePrefix={arXiv},
      primaryClass={quant-ph}, 
}

@misc{nelson2025smallinterestingarchitectureearly,
      title={A small and interesting architecture for early fault-tolerant quantum computers}, 
      author={Jacob S. Nelson and Andrew J. Landahl and Andrew D. Baczewski},
      year={2025},
      eprint={2507.20387},
      archivePrefix={arXiv},
      primaryClass={quant-ph},
}

@misc{koch2025guppypythonicquantumclassicalprogramming,
      title={GUPPY: Pythonic Quantum-Classical Programming}, 
      author={Mark Koch and Alan Lawrence and Kartik Singhal and Seyon Sivarajah and Ross Duncan},
      year={2025},
      eprint={2510.12582},
      archivePrefix={arXiv},
      primaryClass={cs.PL}, 
}

@misc{gidney2025factor,
    title={How to factor 2048 bit RSA integers with less than a million noisy qubits}, 
    author={Craig Gidney},
    year={2025},
    eprint={2505.15917},
    archivePrefix={arXiv},
    primaryClass={quant-ph},
    eprint={2505.15917}, 
}

@article{Babbush_2018,
   title={Encoding Electronic Spectra in Quantum Circuits with Linear T Complexity},
   volume={8},
   ISSN={2160-3308},
   eprint={1805.03662},
   DOI={10.1103/physrevx.8.041015},
   number={4},
   journal={Physical Review X},
   publisher={American Physical Society (APS)},
   author={Babbush, Ryan and Gidney, Craig and Berry, Dominic W. and Wiebe, Nathan and McClean, Jarrod and Paler, Alexandru and Fowler, Austin and Neven, Hartmut},
   year={2018},
   month=Oct }

@article{Wang_2024,
   title={Fault-tolerant one-bit addition with the smallest interesting color code},
   volume={10},
   ISSN={2375-2548},
   eprint={2309.09893},
   DOI={10.1126/sciadv.ado9024},
   number={29},
   journal={Science Advances},
   publisher={American Association for the Advancement of Science (AAAS)},
   author={Wang, Yang and Simsek, Selwyn and Gatterman, Thomas M. and Gerber, Justin A. and Gilmore, Kevin and Gresh, Dan and Hewitt, Nathan and Horst, Chandler V. and Matheny, Mitchell and Mengle, Tanner and Neyenhuis, Brian and Criger, Ben},
   year={2024}}

@misc{koh2026entanglinglogicalqubitsphysical,
      title={Entangling logical qubits without physical operations}, 
      author={Jin Ming Koh and Anqi Gong and Andrei C. Diaconu and Daniel Bochen Tan and Alexandra A. Geim and Michael J. Gullans and Norman Y. Yao and Mikhail D. Lukin and Shayan Majidy},
      year={2026},
      eprint={2601.20927},
      archivePrefix={arXiv},
      primaryClass={quant-ph},
      url={https://arxiv.org/abs/2601.20927}, 
}

@article{heussen2025efficient,
   title={Efficient fault-tolerant code switching via one-way transversal CNOT gates},
   volume={9},
   ISSN={2521-327X},
   url={http://dx.doi.org/10.22331/q-2025-09-03-1846},
   DOI={10.22331/q-2025-09-03-1846},
   journal={Quantum},
   publisher={Verein zur Forderung des Open Access Publizierens in den Quantenwissenschaften},
   author={Heußen, Sascha and Hilder, Janine},
   year={2025},
   pages={1846} 
}

@article{ransford202698,
  title={A 98-qubit trapped-ion quantum computer with all-to-all connectivity},
  author={Ransford, Anthony and Allman, MS and Arkinstall, Jake and Campora, JP and Cooper, Samuel F and Delaney, Robert D and Dreiling, Joan M and Estey, Brian and Figgatt, Caroline and Hall, Alex and others},
  journal={Nature},
  volume={655},
  number={8121},
  pages={81--86},
  year={2026},
  publisher={Nature Publishing Group},
  doi = {10.1038/s41586-026-10676-4},
  eprint = {2511.05465}
}

@article{ryan2021realization,
  title = {Realization of Real-Time Fault-Tolerant Quantum Error Correction},
  author = {Ryan-Anderson, C. and Bohnet, J. G. and Lee, K. and Gresh, D. and Hankin, A. and Gaebler, J. P. and Francois, D. and Chernoguzov, A. and Lucchetti, D. and Brown, N. C. and Gatterman, T. M. and Halit, S. K. and Gilmore, K. and Gerber, J. A. and Neyenhuis, B. and Hayes, D. and Stutz, R. P.},
  journal = {Phys. Rev. X},
  volume = {11},
  issue = {4},
  pages = {041058},
  numpages = {29},
  year = {2021},
  month = {Dec},
  publisher = {American Physical Society},
  doi = {10.1103/PhysRevX.11.041058},
  eprint = {2107.07505}
}

@article{acharya2024quantumerrorcorrectionsurface,
   title={Suppressing quantum errors by scaling a surface code logical qubit},
   volume={614},
   ISSN={1476-4687},
   eprint={2207.06431},
   DOI={10.1038/s41586-022-05434-1},
   number={7949},
   journal={Nature},
   publisher={Springer Science and Business Media LLC},
   author={{Google Quantum AI.}},
   year={2023},
   month=feb, pages={676–681} }

@article{Bluvstein_2023,
   title={Logical quantum processor based on reconfigurable atom arrays},
   volume={626},
   ISSN={1476-4687},
   eprint={2312.03982},
   DOI={10.1038/s41586-023-06927-3},
   number={7997},
   journal={Nature},
   publisher={Springer Science and Business Media LLC},
   author={Bluvstein, Dolev and Evered, Simon J. and Geim, Alexandra A. and Li, Sophie H. and Zhou, Hengyun and Manovitz, Tom and Ebadi, Sepehr and Cain, Madelyn and Kalinowski, Marcin and Hangleiter, Dominik and Bonilla Ataides, J. Pablo and Maskara, Nishad and Cong, Iris and Gao, Xun and Sales Rodriguez, Pedro and Karolyshyn, Thomas and Semeghini, Giulia and Gullans, Michael J. and Greiner, Markus and Vuletić, Vladan and Lukin, Mikhail D.},
   year={2023},
   month=dec, pages={58–65} }

@article{pogorelov2025experimental,
  title={Experimental fault-tolerant code switching},
  author={Pogorelov, Ivan and Butt, Friederike and Postler, Lukas and Marciniak, Christian D and Schindler, Philipp and M{\"u}ller, Markus and Monz, Thomas},
  journal={Nature Physics},
  volume={21},
  number={2},
  pages={298--303},
  year={2025},
  publisher={Nature Publishing Group UK London},
  DOI={10.1038/s41567-024-02727-2},
  eprint={2403.13732}
}

@article{ryan2024high,
  title={High-fidelity teleportation of a logical qubit using transversal gates and lattice surgery},
  author={Ryan-Anderson, Ciaran and Brown, NC and Baldwin, CH and Dreiling, JM and Foltz, C and Gaebler, JP and Gatterman, TM and Hewitt, N and Holliman, C and Horst, CV and others},
  journal={Science},
  volume={385},
  number={6715},
  pages={1327--1331},
  year={2024},
  publisher={American Association for the Advancement of Science},
  eprint={2404.16728},
  DOI={10.1126/science.adp6016}
}

@article{reichardt2024logical,
  title={Logical computation demonstrated with a neutral atom quantum processor},
  author={Reichardt, Ben W and Paetznick, Adam and Aasen, David and Basov, Ivan and Bello-Rivas, Juan M and Bonderson, Parsa and Chao, Rui and van Dam, Wim and Hastings, Matthew B and Paz, Andres and others},
  journal={arXiv preprint arXiv:2411.11822},
  volume={10},
  year={2024},
  eprint={2411.11822}
}

@article{lacroix2025scaling,
  title={Scaling and logic in the colour code on a superconducting quantum processor},
  author={Lacroix, Nathan and Bourassa, Alexandre and Heras, Francisco JH and Zhang, Lei M and Bausch, Johannes and Senior, Andrew W and Edlich, Thomas and Shutty, Noah and Sivak, Volodymyr and Bengtsson, Andreas and others},
  journal={Nature},
  volume={645},
  number={8081},
  pages={614--619},
  year={2025},
  publisher={Nature Publishing Group UK London},
  DOI={10.1038/s41586-025-09061-4},
  eprint={2412.14256}
}

@article{eickbusch2025demonstration,
  title={Demonstration of dynamic surface codes},
  author={Eickbusch, Alec and McEwen, Matt and Sivak, Volodymyr and Bourassa, Alexandre and Atalaya, Juan and Claes, Jahan and Kafri, Dvir and Gidney, Craig and Warren, Christopher W and Gross, Jonathan and others},
  journal={Nature Physics},
  volume={21},
  number={12},
  pages={1994--2001},
  year={2025},
  publisher={Nature Publishing Group UK London},
  eprint={2412.14360},
  DOI={10.1038/s41567-025-03070-w}
}

@article{dasu2025breaking,
  title={Breaking even with magic: demonstration of a high-fidelity logical non-Clifford gate},
  author={Dasu, Shival and Burton, Simon and Mayer, Karl and Amaro, David and Gerber, Justin A and Gilmore, Kevin and Gresh, Dan and DelVento, Davide and Potter, Andrew C and Hayes, David},
  journal={arXiv preprint arXiv:2506.14688},
  year={2025},
  eprint={2506.14688}
}

@article{rosenfeld2025magic,
  title={Magic state cultivation on a superconducting quantum processor},
  author={Rosenfeld, Emma and Gidney, Craig and Roberts, Gabrielle and Morvan, Alexis and Lacroix, Nathan and Kafri, Dvir and Marshall, Jeffrey and Li, Ming and Sivak, Volodymyr and Abanin, Dmitry and others},
  journal={arXiv preprint},
  year={2025},
  eprint={2512.13908}
}

@article{perlin2026fault,
  title={Fault-tolerant execution of error-corrected quantum algorithms},
  author={Perlin, Michael A and He, Zichang and Armenakas, Anthony Alexiades and Andres-Martinez, Pablo and Hao, Tianyi and Herman, Dylan and Jin, Yuwei and Mayer, Karl and Self, Chris and Amaro, David and others},
  journal={arXiv preprint},
  year={2026},
  eprint={2603.04584}
}

@misc{butt2025demonstrationmeasurementfreeuniversalfaulttolerant,
      title={Demonstration of measurement-free universal fault-tolerant quantum computation}, 
      author={Friederike Butt and Ivan Pogorelov and Robert Freund and Alex Steiner and Marcel Meyer and Thomas Monz and Markus Müller},
      journal = {Nat Commun},
      volume = {17},
      issue = {995},
      year={2026},
      DOI={10.1038/s41467-026-68533-x},
      eprint={2506.22600},
}

@misc{wang2026superconductingsurfacecodeprocessorlatticesurgery,
      title={A superconducting surface-code processor with lattice-surgery logical operations}, 
      author={Yanzhe Wang and Fanhao Shen and Haipeng Xie and Aosai Zhang and Yu Gao and Chuanyu Zhang and Xuhao Zhu and Feitong Jin and Yiren Zou and Ning Wang and Zhengyi Cui and Zehang Bao and Zitian Zhu and Jiarun Zhong and Gongyu Liu and Jia-Nan Yang and Yihang Han and Yiyang He and Jiayuan Shen and Han Wang and Jiahua Huang and Xinrong Zhang and Sailang Zhou and Hang Dong and Jinfeng Deng and Yaozu Wu and Zixuan Song and Hekang Li and Zhen Wang and Chao Song and Qiujiang Guo and Pengfei Zhang and H. Wang and Ying Li},
      year={2026},
      eprint={2606.06598},
      archivePrefix={arXiv},
      primaryClass={quant-ph}
}

@article{PRXQuantum.1.010302,
  title = {Flag Fault-Tolerant Error Correction for any Stabilizer Code},
  author = {Chao, Rui and Reichardt, Ben W.},
  journal = {PRX Quantum},
  volume = {1},
  issue = {1},
  pages = {010302},
  numpages = {6},
  year = {2020},
  month = {Sep},
  publisher = {American Physical Society},
  doi = {10.1103/PRXQuantum.1.010302},
  eprint={1912.09549}
}

@misc{gottesman2016quantumfaulttolerancesmall,
      title={Quantum fault tolerance in small experiments}, 
      author={Daniel Gottesman},
      year={2016},
      eprint={1610.03507},
      archivePrefix={arXiv},
      primaryClass={quant-ph},
}

@article{Calderbank_1996,
   title={Good quantum error-correcting codes exist},
   volume={54},
   ISSN={1094-1622},
   eprint={quant-ph/9512032},
   DOI={10.1103/physreva.54.1098},
   number={2},
   journal={Physical Review A},
   publisher={American Physical Society (APS)},
   author={Calderbank, A. R. and Shor, Peter W.},
   year={1996},
   month=Aug, pages={1098–1105} }

@article{steane1996,
   title={Multiple Particle Interference and Quantum Error Correction},
   volume={452},
   ISSN={1471-2946},
   eprint={quant-ph/901029v3},
   DOI={10.1098/rspa.1996.0136},
   number={1954},
   journal={Proceedings of the Royal Society of London. Series A: Mathematical, Physical and Engineering Sciences},
   author={Andrew Steane},
   publisher={The Royal Society},
   year={1996},
   month=Nov, pages={2551–2577} }

@article{rubin2024quantum,
  title={Quantum computation of stopping power for inertial fusion target design},
  author={Rubin, Nicholas C and Berry, Dominic W and Kononov, Alina and Malone, Fionn D and Khattar, Tanuj and White, Alec and Lee, Joonho and Neven, Hartmut and Babbush, Ryan and Baczewski, Andrew D},
  journal={Proceedings of the National Academy of Sciences},
  volume={121},
  number={23},
  pages={e2317772121},
  year={2024},
  doi={10.1073/pnas.2317772121},
  publisher={National Academy of Sciences},
  eprint={2308.12352}
}

@article{pathak2026approach,
  title={An approach for calculating astrophysical opacities on quantum computers},
  author={Pathak, Shivesh and Kononov, Alina and Baczewski, Andrew D},
  journal={arXiv preprint},
  eprint={82607.02811},
  year={2026}
}

@article{berthusen2026experimental,
  title={Experimental validation of a compact fault-tolerant architecture for trapped ions},
  author={Berthusen, Noah and Lavasani, Ali and Benhemou, Asmae and Allman, MS and Dreiling, Joan and Estey, Brian and Foltz, Cameron and Jacobs, Trent and Mills, Michael and Park, Annie Jihyun and others},
  journal={arXiv preprint},
  eprint={2609.03194},
  year={2026}
}

@article{gidney2018halving,
  title={Halving the cost of quantum addition},
  author={Gidney, Craig},
  journal={Quantum},
  volume={2},
  pages={74},
  year={2018},
  doi={10.22331/q-2018-06-18-74},
  publisher={Verein zur F{\"o}rderung des Open Access Publizierens in den Quantenwissenschaften},
  eprint={1709.06648}
}

@misc{preskill2025beyond,
  title={Beyond {NISQ}: The megaquop machine},
  author={Preskill, John},
  journal={ACM Transactions on Quantum Computing},
  volume={6},
  number={3},
  pages={1--7},
  year={2025},
  doi={10.1145/3723153},
  publisher={ACM New York, NY},
  eprint = {2502.17368}
}

@misc{BerlinPaper,
      title={Hardware-tailored logical Clifford circuits for stabilizer codes}, 
      author={Eric J. Kuehnke and Kyano Levi and Joschka Roffe and Jens Eisert and Daniel Miller},
      year={2025},
      eprint={2505.20261},
      archivePrefix={arXiv},
      primaryClass={quant-ph},
      eprint={2505.20261}, 
}

\end{document}